# Symmetry-Dependent Mechanical and Vibrational Response of Formamidinium Lead Halide Perovskites: A DFT Study

Mahdi Faghihnasiri[a,*], Carmine Autieri[b,†] and Sara Memarzadeh[b,c,‡]
[a] Computational Laboratory, Carbon Tech Industrial Group, CarbonTech, Iran
[b] International Research Centre Magtop, Institute of Physics, Polish Academy of Sciences, Aleja Lotników 32/46, 02668 Warsaw, Poland
[c] Institute of Spintronics and Quantum Information, Faculty of Physics and Astronomy, Adam Mickiewicz University Poznań, Uniwersytetu Poznańskiego 2, 61-614 Poznań, Poland

## Abstract

Formamidinium-based hybrid halide perovskites ($FAPbX_3$, X = Cl, Br, and I) have attracted considerable attention for optoelectronic applications owing to their outstanding optical and electronic properties. However, the influence of crystal symmetry reduction on their mechanical behavior and stability has not yet been comprehensively understood. In this work, density functional theory (DFT) calculations were performed to investigate the structural, elastic, dynamical, and nonlinear mechanical properties of the cubic and ps-cubic phases of $FAPbX_3$. The elastic constants, bulk, shear, and Young's moduli, Poisson's ratio, sound velocities, and Debye temperature were evaluated and correlated with the second Piola–Kirchhoff stress–strain response under tensile and compressive loading. The results reveal that the effect of symmetry reduction is strongly dependent on the halide composition. For $FAPbCl_3$ and $FAPbBr_3$, the transition from the cubic to the ps-cubic phase reduces the lattice stiffness, decreases the acoustic phonon velocities, and lowers the Debye temperature, whereas the opposite trend is observed for $FAPbI_3$, where the increased elastic moduli and Debye temperature indicate enhanced lattice rigidity. The stress–strain analysis further reveals pronounced nonlinear, anisotropic, and asymmetric mechanical behavior, demonstrating that symmetry reduction can either activate or suppress strain-accommodation mechanisms depending on the halide species, thereby governing the mechanical stability and the onset of structural softening. These findings provide microscopic insight into the relationship between crystal symmetry, lattice dynamics, and nonlinear mechanical response in formamidinium-based halide perovskites, offering useful guidance for the design of mechanically robust optoelectronic materials.

* mahdi.faghihnasiri@gmail.com
† autieri@magtop.ifpan.edu.pl
‡ saramemarzadeh@magtop.ifpan.edu.pl

## Introduction

The widespread availability of sunlight as an inexhaustible energy source, together with the global warming crisis and the concept of net-zero carbon emissions, has motivated researchers to make extensive efforts toward improving the performance of solar cells. Materials used as light absorbers in solar cells should possess characteristics such as high charge-carrier mobility, a suitable band gap for maximum light absorption, and high efficiency in the separation of electron–hole pairs (1-3). Organic–inorganic hybrid halide perovskite materials have attracted considerable attention in recent years as absorber materials for solar cells due to their favorable properties, including an appropriate optical band gap (around 1.5 eV), high optical absorption coefficient, high charge-carrier mobility, long carrier lifetime, and long diffusion length (4). The new generation of solar cells, known as perovskite solar cells, has gained significant potential for large-scale commercialization in recent years due to its high efficiency, low fabrication cost, and ease of manufacturing (5-7). At present, certain designs of perovskite solar cells have successfully achieved power conversion efficiencies (PCE) exceeding 26% (8-10).

The chemical formula of perovskites in perovskite solar cells is typically $ABX_3$, where A represents a monovalent organic cation [$CH_3NH_3$ (MA), $HC(NH_2)_2$ (FA), Cs], B represents a divalent metal cation ($Pb^{2+}$, $Sn^{2+}$, $Ge^{2+}$), and X represents a halide anion (I, Br, Cl, or a mixture of them) (11-13). Initially, methylammonium (MA)-based lead halide perovskites attracted the most attention, and these solar cells achieved remarkable efficiencies. However, the MA cation induces instability at the center of the lead–halide lattice structure, which hindered the operational stability of these solar cells. In this context, the organic ion formamidinium ($CH(NH_2)_2^+$) has been shown to be one of the few ions capable of fitting into the lead–halide lattice and stabilizing the perovskite structure (14). Since formamidinium-based lead halides ($FAPbX_3$) possess electronic and optical band gaps similar to those of $MAPbX_3$ compounds, they have received considerable attention for improving the efficiency and stability of photovoltaic cells (15). To date, metal halide perovskite solar cells based on formamidinium with $Cs^+$ doping have achieved an efficiency of 26.91%, and have maintained 95% of their initial efficiency under continuous operation at 85 °C for 1500 hours (16).

Although the efficiency of perovskite solar cells has experienced remarkable growth in recent years, it can be confidently stated that the mechanical stability of the absorber layer remains the most important challenge at present (17). Factors such as temperature variations, mismatch between the absorber layer and the charge transport layers, or even the electrodes, subject the perovskite structure to tensile or bending stresses. These stresses can lead to crack formation, interfacial delamination between layers, and ultimately performance degradation and reduced solar cell lifetime (18). Consequently, the elastic behavior and mechanical properties of the perovskite layer play a crucial role in the performance and lifetime of these solar cells (19, 20).

In addition, the development of flexible solar cells as well as wearable electronic devices requires a precise and comprehensive understanding of the mechanical properties of these materials (21). To gain a better understanding of these properties, parameters such as Young's modulus, bulk modulus, and shear modulus can provide a clear picture of the elastic characteristics of these structures (22). Therefore, the investigation of the mechanical properties of these materials is as important as the study of their electronic and optical properties.

Unlike conventional semiconductors, organic–inorganic perovskites possess a relatively soft and deformable lattice. For example, the Young's modulus of single-crystal silicon is approximately 169 GPa along the ⟨111⟩ direction and about 130 GPa along the ⟨100⟩ direction, while a value close to 85 GPa has been reported for GaAs (23). In contrast, nanoindentation measurements on $FAPbBr_3$ and $FAPbI_3$ single crystals have shown that the Young's modulus of these materials lies only within the range of 9.7–12.3 GPa, and their hardness ranges from 0.36 to 0.45 GPa (24).

Guo and co-workers (25) investigated the mechanical properties of formamidinium-based perovskites using the DFT approach. Their calculations yielded a bulk modulus of 18.34 GPa, a shear modulus of approximately 6.5 GPa, and a Young's modulus close to 18 GPa for $FAPbBr_3$. In contrast, for $FAPbI_3$, these values decreased to about 14–15 GPa for the bulk modulus, nearly 5 GPa for the shear modulus, and around 14–15 GPa for the Young's modulus, respectively. The good agreement between the calculated bulk modulus of $FAPbBr_3$ (18.34 GPa) and the experimental value (16.9 GPa) demonstrates the effectiveness of the DFT approach for calculating the mechanical properties of perovskites.

The crystal structure of $FAPbX_3$ (X = Cl, Br, and I) is strongly temperature-dependent and undergoes phase transitions to different crystal phases with changing temperature (26). These materials are generally stable in the cubic phase, which possesses the highest symmetry, at high temperatures, particularly at room temperature (27). As the temperature decreases, they transform into lower-symmetry phases such as tetragonal and orthorhombic structures due to the tilting of $PbX_6$ octahedra as well as changes in the orientation and ordering of the organic $FA^+$ cations. These phase transitions play an important role in determining the structural stability, lattice interactions, and physical properties of the materials, and can significantly affect the electronic, optical, and mechanical characteristics of $FAPbX_3$ perovskites. However, since these structures are generally in the cubic phase under operating conditions, the present study focuses on the investigation of the cubic phase with space group $Pm\bar{3}m$.

Although extensive studies have been conducted on the electronic and optical properties of perovskites, systematic investigations of the elastic properties and mechanical stability of these structures remain limited. A comprehensive evaluation of these properties requires the calculation of elastic constants as the fundamental parameters governing mechanical behavior. To calculate these constants as accurately as possible, the density functional theory (DFT) approach has been employed in the present study. Furthermore, the effect of halide substitution on these properties has been investigated through the examination of $FAPbCl_3$, $FAPbBr_3$, and $FAPbI_3$. The energy-strain curves were obtained and fitted using second order polynomial functions. We applied Lagrangian strains to all structures and calculated the elastic constants. Then we calculated mechanical properties like the Young's, bulk, and shear moduli.

## Model and Computational Methods

### DFT calculations

All the energies and forces were calculated using density functional theory (DFT) with the Quantum ESPRESSO package (28) to study the mechanical properties of $FAPbX_3$ (X=Cl, Br, and I) compounds. The generalized gradient approximation (GGA) with the Perdew-Burke-Ernzerhof for solids (PBEsol) (29) correction was used for the exchange-correlation function. To achieve higher accuracy in describing the electronic structure and electron–nucleus interactions, Projector Augmented Wave (PAW) pseudopotentials were employed in all calculations. The k-point mesh was converged using a 12×12×12 Monkhorst-Pack grid, and a plane wave basis set with a 650 eV cut-off energy was chosen. All compounds were geometrically optimized using the Broyden-Fletcher-Goldfarb-Shanno (BFGS) algorithm (30) until the total energy converged to less than 1 meV/atom and the force on each atom was less than 1 meV/Å. **Figure 1** shows the unit cell of $FAPbX_3$ (X= Cl, Br, and I).

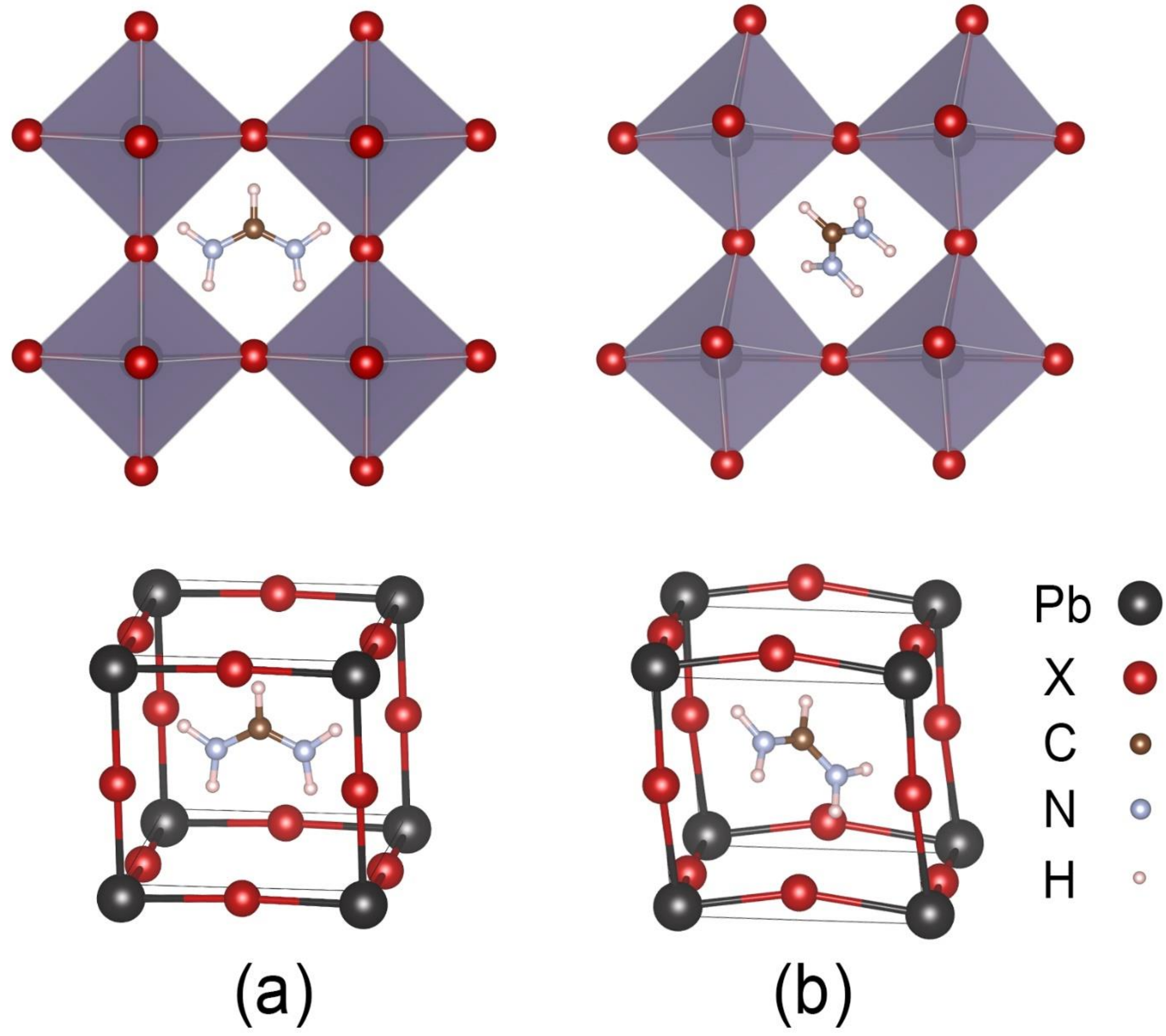

**Figure 1.** Crystal structures of $FAPbX_3$ perovskites (X = Cl, Br, and I) in the (a) cubic and (b) pseudo-cubic phases. The upper panels show the corner-sharing $PbX_6$ octahedral framework, while the lower panels present the atomic representation of the unit cell. Pb and X atoms are represented by grey and red spheres, respectively. The FA cations are depicted with carbon, nitrogen, and hydrogen atoms shown as brown, light blue, and white spheres, respectively.

**Elastic constants calculation**

Continuum mechanics methods (31) were employed to evaluate the elastic constants of the perovskite compounds. The total energy of the system and the strain have the following relationship:

$$E(\eta_I) = E(0) + V_0 \sum_{I=1}^{3} \tau_I^{(0)} \eta_I + \frac{V_0}{2} \sum_{I,J=1}^{3} C_{IJ} \eta_I \eta_J \tag{1}$$

The energy and volume in the equilibrium state are denoted by $E(0)$ and $V_0$, while η and $\tau$ are the Lagrangian strains and stress tensor, respectively. $C$ represents the elastic constants of the structure. The elastic constants can be calculated based on Eq (1) as follows (using the Voigt notation (32)):

$$C_{IJ} = \frac{1}{V_0} \frac{\partial^2 E}{\partial \eta_I \partial \eta_J} \tag{2}$$

For cubic structures, only three independent elastic constants, $C_{11}$, $C_{12}$, and $C_{44}$, are required to fully describe the elastic response (Eq (2)). In the present work, the optimized pseudo-cubic structures were treated within the cubic approximation for the evaluation of the elastic constants and mechanical stability. The elastic constants were calculated based on the energy-strain relations described by Jamal et al. (33). They defined three different deformation tensors to extract the energy-strain relations:

$$D_1 = \begin{bmatrix} \eta & 0 & 0 \\ 0 & \eta & 0 \\ 0 & 0 & \eta \end{bmatrix}$$

$$D_2 = \begin{bmatrix} \eta & 0 & 0 \\ 0 & -\eta & 0 \\ 0 & 0 & \frac{\eta^2}{1-\eta^2} \end{bmatrix} \tag{3}$$

$$D_3 = \begin{bmatrix} 0 & \eta & 0 \\ \eta & 0 & 0 \\ 0 & 0 & \frac{\eta^2}{1-\eta^2} \end{bmatrix}$$

The deformation ranges are set from -10% to 20% for these structures. For the calculation of the second-order elastic constants, a strain range of −2% to +2% was employed to ensure the validity of Hooke's law and remain within the harmonic elastic regime. Strain values beyond ±2% were used to investigate nonlinear behavior and to examine possible phase transitions. strain cannot exceed certain limits without violating the physical laws or causing irreversible damage to the material. According to the Lennard-Jones potential (34), the repulsive force is stronger in compressive strain compared to tensile strain. Therefore, a compressive strain of -10% is sufficient. Excessive negative strain lacks physical meaning as the material cannot realistically be compressed to such an extent. Similarly, excessive tensile strain does not provide valuable information for studying the elastic properties of the material as it leads to material failure.

**Mechanical properties**

The calculated elastic constants were further employed to evaluate the macroscopic mechanical properties of the investigated $FAPbX_3$ compounds. The bulk modulus (B), shear modulus (G), Young's modulus (E), and Poisson's ratio (ν) can be calculated using the obtained elastic constants. The determination of B and G moduli can be approached using various theories, but we have chosen to utilize the Voigt theory (35). Since the focus of this study is on comparing behavioral trends among compounds with similar structures, the Voigt approximation provides an acceptable and reasonable estimate of the effective elastic moduli. In the Voigt theory, for the cubic structure, the expressions for B and G are as follows:

$$B = \frac{C_{11} + 2C_{12}}{3} \quad (4)$$

$$G = \frac{C_{11} - C_{12} + 3C_{44}}{5} \quad (5)$$

The Young's modulus and Poisson's ratio can be obtained by using B and G as follows:

$$E = \frac{9BG}{3B + G} \quad (6)$$

$$\nu = \frac{3B - 2G}{2(3B + G)} \quad (7)$$

**Debye temperature and sound velocity**

To further investigate the bonding strength and lattice stiffness of the studied perovskites, the Debye temperature ($\theta_D$) was estimated using the calculated elastic properties. $\theta_D$ has been calculated using mechanical characteristics such as the bulk modulus, shear modulus, and elastic constants, to assess the strength of the average chemical bonds in these materials. The Debye temperature is often correlated with other physical properties such as the melting point and specific heat. The Debye temperature can be represented by using the average sound velocity ($\nu_m$) that is expressed as follows:

$$\theta_D = \frac{h}{k}\left[\frac{3n}{4\pi}\left(\frac{N_A\rho}{M}\right)\right]^{1/3} \nu_m \quad (8)$$

Here, h, k, n, $N_A$, ρ, M denote Planck's constant, Boltzmann's constant, number of atoms, Avogadro's number, density and molecular mass, respectively. The shear and compressional velocities ($\nu_t$ and $\nu_l$), related to the average sound velocity ($\nu_m$) and can be defined as follows:

$$\nu_t = \sqrt{G/\rho} \quad (9)$$

$$\nu_l = \sqrt{\left(B + \frac{4}{3}G\right)/\rho} \quad (10)$$

$$\nu_m = \left[\frac{1}{3}\left(\frac{2}{\nu_l{}^3} + \frac{1}{\nu_t{}^3}\right)\right]^{-1/3} \quad (11)$$

**Stress-strain relationship**

Beyond the small-strain regime used for determining the second-order elastic constants, the nonlinear mechanical response of the investigated compounds was examined through stress-strain analysis. The stress versus Lagrangian strain relationship has been studied using the second Piola-Kirchhoff (PK2) method, considering Cauchy stresses (36). The PK2 and Cauchy stresses are interconnected as follows:

$$\Sigma = J.\, F^{-1}\sigma(F^{-1})^T \quad (12)$$

F is the deformation gradient tensor, J is the determinant of F, and σ represents Cauchy stresses.

**Results and discussions**

**Structural properties**

According to previous literature reports, $FAPbX_3$ compounds exhibit cubic phases at room temperature, where all lattice vectors are perpendicular to each other (12). However, geometric optimization by variable cell calculation reveals a slight change in the angles between lattice vectors (**Table 1**), referred to as a pseudo-cubic (ps-cubic) structure. Although the lattice vector lengths remain largely unchanged, the lattice angles deviate from 90° by approximately 3-9°, indicating a slight symmetry distortion from the ideal cubic structure and the formation of a pseudo-cubic (ps-cubic) phase. This deviation is attributed to the orientation of the FA cation, which is governed by hydrogen-bonding interactions between the amino groups of the FA cation and the halogen atoms within the $PbX_6$ framework (37, 38). Since the size of the FA molecule is comparable to the dimensions of the cavities in the perovskite lattice, the orientation and rotation of this cation within the unit cell can alter the distribution of internal lattice forces and lead to slight deviations of the lattice angles from the ideal values of a cubic structure. The end-to-end distance between H atoms in the FA cation measures 4.03 Å, which is comparable to the length of lattice vectors, allowing the FA cation to rotate and preferentially align between opposite corners of the cubic unit cell. This orientation contributes to the slight variation in lattice angles. The influence of FA orientation on the structural symmetry of $FAPbI_3$ has also been reported previously (39, 40). As shown in **Table 1**, the values of α and β show an increasing trend with the larger radius of the halogen atoms and the expansion of the unit cell volume. Previous studies have also reported the influence of halide ions on the orientation of FA cations, as well as the resulting changes in the structural parameters and phase transitions of FAPbX3 perovskites (12). The FA cation is not arranged completely randomly even in the average cubic phase; rather, it continuously reorients among several preferred directions and, through N–H···X hydrogen bonding, induces local distortions in the $PbX_6$ octahedra. Although these distortions may appear as an average cubic structure in diffraction patterns, at the atomic scale they lead to local symmetry breaking and the formation of a ps-cubic structure (41).

**Table 1.** Optimized lattice parameters (a, b, and c in Å) and lattice angles (α, β, and γ in degrees) for the $FAPbX_3$ (X = Cl, Br, and I) structures. For comparison, the last column includes the reported lattice constants from the literature.

| Sample | | a (Å) | b (Å) | c (Å) | α | β | γ | Comparison |
|---|---|---|---|---|---|---|---|---|
| $FAPbCl_3$ | cubic | 5.75 | 5.75 | 5.75 | 90 | 90 | 90 | 5.62 (42), 5.65 (43) |
| | ps-cubic | 5.73 | 5.72 | 5.65 | 81.31 | 83.81 | 86.28 | |
| $FAPbBr_3$ | cubic | 5.99 | 5.99 | 5.99 | 90 | 90 | 90 | 5.77 (42), 5.93 (12),5.94 (25) |
| | ps-cubic | 5.97 | 5.96 | 5.09 | 82.38 | 84.51 | 86.48 | |

| $FAPbI_3$ | cubic | 6.40 | 6.25 | 6.33 | 90 | 90 | 90 | 6.36 (39), 6.33 (12), 6.40 (44), |
|---|---|---|---|---|---|---|---|---|
| | ps-cubic | 6.33 | 6.31 | 6.27 | 83.80 | 85.57 | 86.58 | |

**Elastic properties**

The elastic response of the $FAPbX_3$ compounds was analyzed using the calculated energy-strain relationships. **Figure 2** presents the corresponding energy-strain curves obtained for the three deformation tensors. In all compounds, the energy-strain graphs exhibit a parabolic behavior within the range of -2% < η < +2% around the free strain state (0% strain), which is referred to as the harmonic region.-In **Figure 2**, the energy is plotted relative to the equilibrium energy $E_0$, where $E_0$corresponds to the energy of the free-strain (equilibrium) state. Within the harmonic region, a change in the volume of the compounds under tensile or compressive strain leads to a deviation from the stable energy level. The three independent elastic constants extracted from the energy-strain curves are summarized in **Table 2**.

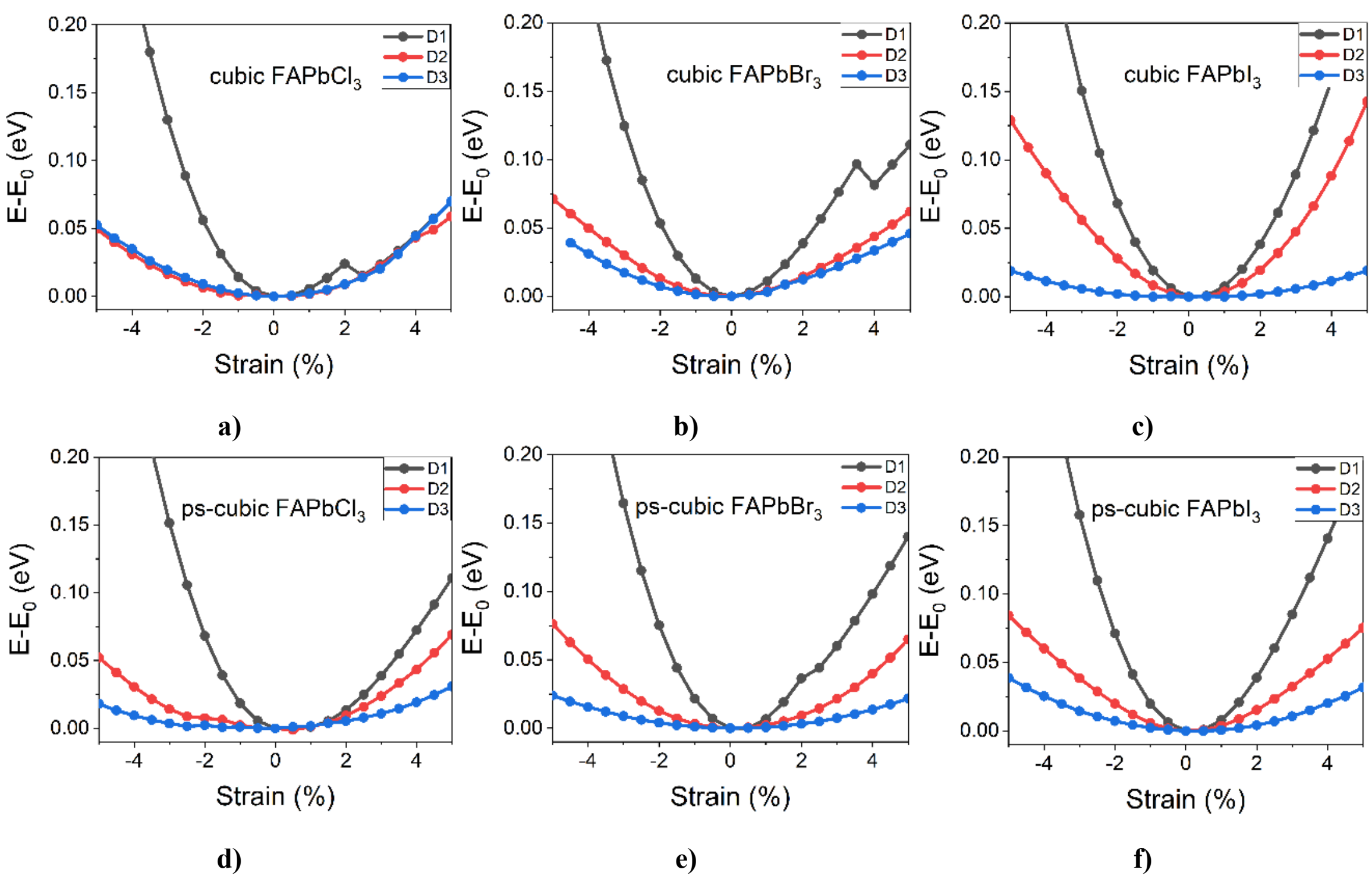

**Figure 2.** Energy–strain curves of $FAPbX_3$ structures under three deformation tensors ($D_1$, $D_2$, and $D_3$) for (a, d) X = Cl, (b, e) X = Br, and (c, f) X = I. The upper and lower panels correspond to the cubic and ps-cubic structures, respectively.

**Table 2**. Calculated elastic constants $C_{ij}$ for the cubic and ps-cubic phases of $FAPbX_3$ (X = Cl, Br, and I) compounds.

| Sample | | $C_{11}$ (GPa) | $C_{12}$ (GPa) | $C_{44}$ (GPa) |
|---|---|---|---|---|
| $FAPbCl_3$ | cubic | 30.25 | 13.02 | 9.68 |
| | ps-cubic | 30.45 | 10.3 | 3.84 |
| $FAPbBr_3$ | cubic | 33.24 | 7.83 | 7.21 |
| | ps-cubic | 33.11 | 11.46 | 5.18 |
| $FAPbI_3$ | cubic | 28.65 | 4.07 | 1.56 |
| | ps-cubic | 37.73 | 9.99 | 4.54 |

The mechanical stability of the cubic and ps-cubic $FAPbX_3$ (X = Cl, Br, and I) compounds was evaluated using the Born-Huang elastic stability criteria for cubic systems (45). As summarized in **Table 2**, all calculated elastic constants satisfy these stability criteria, indicating that both the cubic and ps-cubic $FAPbX_3$ (X = Cl, Br, and I) structures are mechanically stable. The values of $C_{11}$ in the cubic and ps-cubic forms of $FAPbCl_3$ and $FAPbBr_3$ structures are relatively close. However, in the ps-cubic form of the $FAPbI_3$ structure, $C_{11}$ is approximately 9 GPa higher compared to its cubic form. This difference is associated with the symmetry distortion induced by the orientation of the FA cation within the Pb-I framework.

The calculated elastic constants (**Table 2**) show that the Born criteria are satisfied for all cubic and ps-cubic $FAPbX_3$ structures, indicating that they do not exhibit intrinsic mechanical instability from the viewpoint of elasticity. The $C_{11}$ elastic constant for cubic $FAPbBr_3$ was calculated to be 33.24 GPa, which is in close agreement with the previously reported computational value of 34 GPa (46). In addition, this result shows very good agreement with experimental measurements based on Brillouin scattering, which reported a value of approximately 31 GPa by Ferreira et al. (47). Similarly, for α-$FAPbI_3$, the calculated $C_{11}$ value was 28.65 GPa, which differs only slightly from the experimentally reported value of approximately 27 GPa (47).

Overall, for each of these compounds, $C_{11}$ is the largest component of the elastic tensor, whereas $C_{44}$ is the smallest, indicating that the structures are much more resistant to compressive strain than to shear deformation. This behavior suggests that the initial deformation of the perovskite lattice occurs primarily through the rotation and distortion of the $PbX_6$ octahedra rather than through the compression of the Pb–X bonds (48). Among them, cubic $FAPbI_3$ exhibits the smallest $C_{44}$ value (1.56 GPa).

The calculated elastic properties such as B, G, E, ν, and B/G are given in **Table 3**. In cubic structures, an increase in the halide radius leads to an increase in bond lengths and a reduction in orbital overlap, which weakens the B–X bonds and consequently decreases the B, E, and G moduli (49). Specifically, $FAPbCl_3$ exhibits the highest bulk modulus, while $FAPbI_3$ has the lowest bulk modulus, with a significant difference of 6 GPa.

An interesting point is that the decreasing trend of the bulk modulus in the cubic phase with increasing halogen radius is not preserved in the ps-cubic phase and has changed. In ps-cubic perovskites, the bulk modulus increases with the larger halogen radius. Notably, the bulk modulus of ps-cubic $FAPbI_3$ (19.24 GPa) surpasses that of all other structures.

**Table 3**. Calculated bulk modulus (B), Young's modulus (E), shear modulus (G), Poisson's ratio (ν) and B/G for cubic and ps-cubic of $FAPbX_3$ (X= Cl, Br, and I)

| Sample | | B (GPa) | E (GPa) | G (GPa) | ν | B/G |
|---|---|---|---|---|---|---|
| $FAPbCl_3$ | cubic | 18.76 | 23.84 | 9.25 | 0.28 | 2.02 |
| | ps-cubic | 17.02 | 16.9 | 6.33 | 0.33 | 2.68 |
| $FAPbBr_3$ | cubic | 16.3 | 23.67 | 9.41 | 0.25 | 1.73 |
| | ps-cubic | 18.68 | 19.7 | 7.43 | 0.32 | 2.51 |
| $FAPbI_3$ | cubic | 12.27 | 15.15 | 5.85 | 0.29 | 2.09 |
| | ps-cubic | 19.24 | 21.7 | 8.27 | 0.31 | 2.32 |

Furthermore, the larger size of the Pb-I framework in comparison to Pb-Br and Pb-Cl plays a significant role in enhancing the stability of the structure, as it can accommodate the sizable FA structure with a length of 4.03 Å. The diagonal orientation of the FA cation within the Pb-I framework, also measuring 4.03 Å, contributes to greater stability and further ensures structural integrity. Sharma et al. (50) showed that the dynamics and orientation of FA depend on the type of halide. Changing the halide alters the rotational barrier of FA and consequently changes the stress distribution within the perovskite unit cell. In the ideal cubic structure, the bulk modulus is mainly governed by the strength of the Pb–X bonds, whereas in the ps-cubic phase, the reorientation of the FA cation changes the stress distribution within the lattice and alters the halogen-dependent trend. In ps-cubic structures, the reorientation of the FA cation not only changes the lattice angles but also slightly deviates the Pb–X–Pb bond angles from the ideal value of 180°. Hydrogen bonding between the $NH_2$ groups of the FA cation and halide ions also plays a decisive role in the formation of this distortion. These interactions lead to local lattice strain, changes in Pb–X–Pb angles, and a reduction in lattice stiffness along certain crystallographic directions. As a result, even without a noticeable change in the lattice parameters, the elastic response of the material can be significantly altered (51).

The calculated Young's moduli for the cubic phases of $FAPbCl_3$, $FAPbBr_3$, and $FAPbI_3$ are 23.84, 23.67, and 15.15 GPa, respectively, whereas in the ps-cubic phases these values change to 16.90, 19.70, and 21.70 GPa. The Young's modulus of cubic $FAPbBr_3$ is in good agreement with experimental nanoindentation results reported in the range of 20–22 GPa (24). For $FAPbI_3$, values of approximately 14–17 GPa (24) and around 16 GPa from previous DFT calculations (37) have been reported, which are consistent with the present result (15.15 GPa, **Table 3**). Finally, the cubic-phase result for $FAPbCl_3$ (23.84 GPa) indicates a higher stiffness compared to both bromide and iodide counterparts.

In contrast, the ps-cubic phase exhibits a different behavior. In $FAPbCl_3$, the Young's modulus decreases from 23.84 to 16.90 GPa, corresponding to a reduction of approximately 29%, while in $FAPbBr_3$ it decreases from 23.67 to 19.70 GPa, representing a reduction of approximately 17%. This trend is consistent with the structural distortion and stress redistribution discussed above, indicating that the effect of FA reorientation depends strongly on the halide species.

In the cubic structures, the shear modulus increases from 9.25 GPa for $FAPbCl_3$ to 9.41 GPa for $FAPbBr_3$ and then decreases to 5.85 GPa for $FAPbI_3$. This trend is consistent with the results reported by Guo et al. (25), who obtained values of 6.37 GPa for cubic $FAPbBr_3$ and 5.59 GPa for cubic $FAPbI_3$, demonstrating that substituting Br for I enhances the shear resistance of the lattice. Similarly, Mayengbam et al.(46) reported shear modulus values of 6.60 GPa for $FAPbBr_3$ and 3.60 GPa for $FAPbI_3$, further confirming the same trend.

In contrast, the behavior of the ps-cubic phase is not uniform and reflects the direct influence of the structural distortion induced by the orientation of the FA cation on the shear stiffness of the lattice. In $FAPbCl_3$, the shear modulus decreases from 9.25 to 6.33 GPa (approximately a 32% reduction), while in $FAPbBr_3$ it decreases from 9.41 to 7.43 GPa (approximately a 21% reduction).

A different trend is observed for $FAPbI_3$, with the shear modulus increasing from 5.85 GPa in the cubic structure to 8.27 GPa in the pseudocubic structure (approximately a 41% increase). This behavior further demonstrates that the effect of pseudo-cubic distortion on the shear response is strongly dependent on the halide species.

The Pugh criterion (52) indicates that the B/G ratio determines whether a material is ductile or brittle. If this ratio is greater than 1.75, the structure is considered ductile. As shown in **Table 3**, the B/G ratio for all structures is greater than 1.75, indicating that $FAPbX_3$ (X = Cl, Br, and I) perovskites are ductile materials, except for $FAPbBr_3$ in the ps-cubic structure.

**Figure 2** shows the variation of the total energy of the cubic and ps-cubic $FAPbX_3$ (X = Cl, Br, and I) structures as a function of the applied strain for the three independent deformation modes ($D_1$, $D_2$, and $D_3$). For all compounds, the energy curves exhibit an almost ideal parabolic behavior, indicating that the structures remain within the elastic regime and confirming the validity of the energy–strain method for extracting the elastic constants. The most important feature of **Figure 2** is the presence of discontinuities in the $D_1$ energy curves of the cubic $FAPbCl_3$ and $FAPbBr_3$ structures at tensile strains of 2% and 3%, respectively. Since, in the present study, the structure was fully relaxed after the application of each strain value, these discontinuities cannot be attributed to numerical errors or convergence issues. Instead, this behavior indicates that, under the applied strain, the system is transferred from one local minimum on the energy surface to another, resulting in a change in the optimal relaxation pathway. This rearrangement arises from the rotation of the FA molecule, a phenomenon that has been widely reported in hybrid perovskites because of the strong coupling between the inorganic framework and the orientation of the organic cations (12). Recent theoretical studies have also shown that the orientation of the FA cation and its interaction with the inorganic framework can alter the energy difference between nearly degenerate structures by only a few meV, and therefore even small applied strains can change the relative stability of these states.

In the ps-cubic structures, all three compounds exhibit smoother and more continuous behavior than the cubic phase, and no distinct discontinuities are observed in the energy curves. This indicates that the initial distortion present in the ps-cubic structure has already released part of the internal lattice stress before the application of strain. Consequently, when external strain is applied, the structure relaxes gradually rather than undergoing a sudden rearrangement. As a result, the energy surface in this phase is smoother, and the probability of forming new metastable states is reduced. This behavior is consistent with the models proposed for the ordering of the FA cation and its interaction with the inorganic framework in $FAPbX_3$ perovskites, which show that the initial lattice distortion smooths the energy surface and reduces strain-induced structural instabilities (12).

**Figures S1** and **S2** depict three-dimensional (3D) surface contours illustrating the Young's modulus, shear modulus, Poisson's ratio, and linear compressibility of $FAPbCl_3$, $FAPbBr_3$, and $FAPbI_3$ for both cubic (**Figure S1**) and ps-cubic (**Figure S2**) structures. The linear compressibility is derived from the volume compressibility and the bulk modulus, and the ELATE code (53) was employed to generate the 3D surface contours. The shape of the surface graphs reveals fluctuations of the parameters in various directions. An isotropic material would exhibit a spherical three-dimensional dependence, and deviation from this shape indicates anisotropy. The graphs illustrate that linear compressibility exhibits isotropy in all cubic and ps-cubic structures. However, when considering Young's modulus and shear modulus, only the cubic phase of $FAPbCl_3$ demonstrates isotropic behavior, while other structures exhibit anisotropy. Additionally, Poisson's ratio is found to be anisotropic in all structures.

The minimum and maximum values for Young's modulus, shear modulus, Linear compressibility and, Poisson's ratio are tabulated in the **Table 4**. The results highlight the anisotropic nature of these parameters. According to **Table 4**, linear compressibility demonstrates isotropy in all cubic and ps-cubic forms of perovskites, indicating that it is independent of the direction studied. However, the remaining parameters exhibit anisotropy, with their values significantly influenced by the applied force direction.

**Table 4**. Minimum and maximum values of the Young's modulus, linear compressibility, shear modulus, and Poisson's ratio for the cubic and ps-cubic phases of $FAPbBr_3$, $FAPbCl_3$, and $FAPbI_3$.

| Sample | | Young's modulus (GPa) | | Linear compressibility ($TPa^{-1}$) | | Shear modulus (GPa) | | Poisson's ratio | |
|---|---|---|---|---|---|---|---|---|---|
| | | $E_{min}$ | $E_{max}$ | $\beta_{min}$ | $\beta_{max}$ | $G_{min}$ | $G_{max}$ | $\nu_{min}$ | $\nu_{max}$ |
| $FAPbCl_3$ | cubic | 22.41 | 24.78 | 17.76 | 17.76 | 8.61 | 9.68 | 0.24 | 0.32 |
| | ps-cubic | 10.71 | 25.24 | 19.59 | 19.59 | 3.84 | 10.07 | 0.12 | 0.62 |
| $FAPbBr_3$ | cubic | 18.85 | 30.25 | 20.45 | 20.45 | 7.21 | 12.70 | 0.13 | 0.44 |

| | ps-cubic | 14.22 | 27.22 | 17.85 | 17.85 | 5.18 | 10.82 | 0.15 | 0.55 |
|---|---|---|---|---|---|---|---|---|---|
| $FAPbI_3$ | cubic | 4.49 | 27.64 | 27.18 | 27.18 | 1.56 | 12.29 | 0.02 | 0.82 |
| | ps-cubic | 12.63 | 33.55 | 17.33 | 17.33 | 4.54 | 13.87 | 0.09 | 0.64 |

Specifically, for cubic $FAPbCl_3$ perovskite, the maximum and minimum Young's modulus have a small difference of 2 GPa. In contrast, the ps-cubic structure of $FAPbCl_3$ shows a significant difference of approximately 15 GPa, which means the Young's modulus highly sensitive to the investigated direction. Both the cubic and ps-cubic structures of $FAPbBr_3$ exhibit anisotropy, with a notable disparity between the maximum and minimum Young's modulus values. The ps-cubic structure of $FAPbI_3$ displays the highest Young's modulus at 33.55 GPa, while the cubic $FAPbI_3$ exhibits the largest difference of 23 GPa between the maximum and minimum Young's modulus values. These variations signify differing resistance levels to external forces in different directions, with some directions displaying high strength and others displaying weakness and fragility.

Furthermore, **Table 4** reports the maximum and minimum shear modulus values, which highlight the anisotropic nature of both cubic and ps-cubic phases. The cubic structure of $FAPbBr_3$ exhibits the highest shear modulus at 12.70 GPa, whereas the cubic structure of $FAPbI_3$ displays the smallest shear modulus at 1.56 GPa. Similarly, Poisson's ratio demonstrates anisotropic behavior, with varying values observed in different directions.

Debye temperature ($\theta_D$), average sound velocity ($\nu_m$), shear velocity ($\nu_t$), and compressional velocity ($\nu_l$) are provided in **Table 5** for the structures of $FAPbCl_3$, $FAPbBr_3$, and $FAPbI_3$ in both phases. According to the elasticity relations, $\nu_t$ is mainly dependent on the shear modulus (G), whereas $\nu_l$ depends on both the bulk modulus (B) and the shear modulus. Meanwhile, $\nu_m$, and consequently $\theta_D$, are also directly related to these two quantities. Therefore, either a reduction in lattice stiffness or an increase in atomic mass leads to a decrease in the phonon propagation velocities and, ultimately, a reduction in the Debye temperature. This relationship has also been reported in numerous DFT studies on perovskites (54-56).

**Table 5**. The calculated $\nu_t$, $\nu_l$, $\nu_m$, and $\theta_D$ for $FAPbCl_3$, $FAPbBr_3$ and $FAPbI_3$ structures.

| Sample | $\nu_t$ ($^m/_s$) | $\nu_l$ ($^m/_s$) | $\nu_m$ ($^m/_s$) | $\theta_D$ (K) |
|---|---|---|---|---|

| | | | | | |
|---|---|---|---|---|---|
| $FAPbCl_3$ | cubic | 1721.75 | 3156.36 | 2261.07 | 267.87 |
| | ps-cubic | 1446.49 | 2900.29 | 1937.63 | 227.19 |
| $FAPbBr_3$ | cubic | 1577.11 | 2761.34 | 2046.69 | 232.63 |
| | ps-cubic | 1385.31 | 2716.26 | 1847.23 | 211.67 |
| $FAPbI_3$ | cubic | 1465.68 | 2713.99 | 1929.46 | 180.74 |
| | ps-cubic | 1414.90 | 2706.29 | 1876.60 | 201.98 |

In the cubic structures, the trend is completely systematic. With the substitution of Cl by Br and then I, the average sound velocity decreases from 2261.07 m $s^{-1}$ for $FAPbCl_3$ to 2046.69 m $s^{-1}$ for $FAPbBr_3$ and 1929.46 m $s^{-1}$ for $FAPbI_3$. The same trend is observed for $\nu_t$, which decreases from 1721.75 to 1577.11 and then 1465.68 m $s^{-1}$, while $\nu_l$ decreases from 3156.36 to 2761.34 and 2713.99 m $s^{-1}$. This trend is fully consistent with the reduction in the elastic moduli reported in the previous section as well as the increase in the atomic mass of the halogens. Consequently, the Debye temperature decreases from 267.87 K for $FAPbCl_3$ to 232.63 K and then 180.74 K, indicating the gradual softening of the crystal lattice and a reduction in the maximum phonon frequency. Such behavior has previously been reported for hybrid perovskites and other $ABX_3$ families, where the increase in the mass of the X ion leads to lower sound velocities and a lower $\theta_D$ (54).

The behavior of the ps-cubic structures, similar to the results obtained for the Young's and shear moduli, is not completely uniform. In these structures, $FAPbCl_3$ still exhibits the highest average sound velocity (1937.63 m $s^{-1}$) and the highest $\theta_D$ (227.19 K). However, contrary to expectations, $FAPbI_3$ exhibits a higher $\nu_m$ (1876.60 m $s^{-1}$) than $FAPbBr_3$ (1847.23 m $s^{-1}$), while its $\theta_D$ (201.98 K) remains lower than that of $FAPbBr_3$ (211.67 K). This behavior results from the competition between two main factors. On the one hand, the increase in the atomic mass of the halogen tends to reduce the propagation velocity of sound waves. On the other hand, the pseudocubic distortion increases the lattice stiffness in the iodide compound and enhances its elastic moduli relative to $FAPbBr_3$. Therefore, the increase in lattice stiffness in $FAPbI_3$ largely compensates for the effect of the higher atomic mass of iodine, causing the average sound velocity to become nearly equal to, and even slightly higher than, that of $FAPbBr_3$. This trend is also consistent with the elastic behavior discussed above and with previous Brillouin spectroscopy and phonon study, which has demonstrated the influence of FA-cation dynamics and its interaction with the $PbX_6$ framework on the elastic stiffness and acoustic-wave propagation in $FAPbX_3$ perovskites (57).

A comparison between the two phases shows that the transition from the cubic to the ps-cubic structure leads to a significant reduction in all sound velocities and the Debye temperature for $FAPbCl_3$ and $FAPbBr_3$. Specifically, $\theta_D$ decreases from 267.87 to 227.19 K (approximately a 15% reduction) for $FAPbCl_3$ and from 232.63 to 211.67 K (approximately a 9% reduction) for $FAPbBr_3$. This decrease indicates a softening of the lattice following the symmetry distortion. In contrast, $FAPbI_3$ exhibits a different behavior, with $\theta_D$ increasing from 180.74 K in the cubic phase to 201.98 K in the pseudocubic phase (approximately a 12% increase). This result indicates that the pseudocubic distortion in the iodide compound, in contrast to the other two compounds, leads to an increase in lattice stiffness and strengthens the acoustic modes. Since the Debye temperature is proportional to the maximum energy of the acoustic phonons, the increase in $\theta_D$ for $FAPbI_3$ may indicate a reduction in lattice softening and an enhancement of its vibrational stability. This observation is also consistent with the increase in the elastic moduli and the energy–strain behavior of this structure. This relationship among $\theta_D$, the elastic moduli, and phonon stability has also been reported in recent theoretical studies on halide perovskites (57).

**Figure 3** presents the PK2 stress response as a function of strain for the $FAPbX_3$ compounds. For all samples, the stress–strain behavior is clearly nonlinear and asymmetric over the strain range from −10% to +10%, with the magnitude of compressive stress increasing more rapidly than that of tensile stress. This behavior originates from the reduction of interatomic distances under compression, causing the Pb–X bonds and ionic interactions to enter the strongly repulsive region of the interatomic potential, which is physically comparable to the repulsive part of the Lennard–Jones potential. In contrast, under tensile loading, the crystal can partially accommodate the applied strain through elongation of the Pb–X bonds, variation of the Pb–X–Pb bond angles, octahedral tilting, and reorientation of the FA cations. Consequently, the stress increases more gradually under tension, resulting in a more pronounced nonlinear response. The onset of nonlinearity, followed by the attainment of a maximum stress and its subsequent decrease or saturation, indicates that the structures have departed from the linear elastic regime and entered a mechanically unstable state. Since the crystal structure was fully relaxed at each strain level, the discontinuities and abrupt changes in the slope of the stress–strain curves reflect genuine structural rearrangements, including modifications in the octahedral tilting pattern and changes in the orientation of the FA cations. This interpretation is consistent with previous studies highlighting the soft lattice nature of hybrid halide perovskites and the critical roles of octahedral tilting and strain in governing their mechanical response (58).

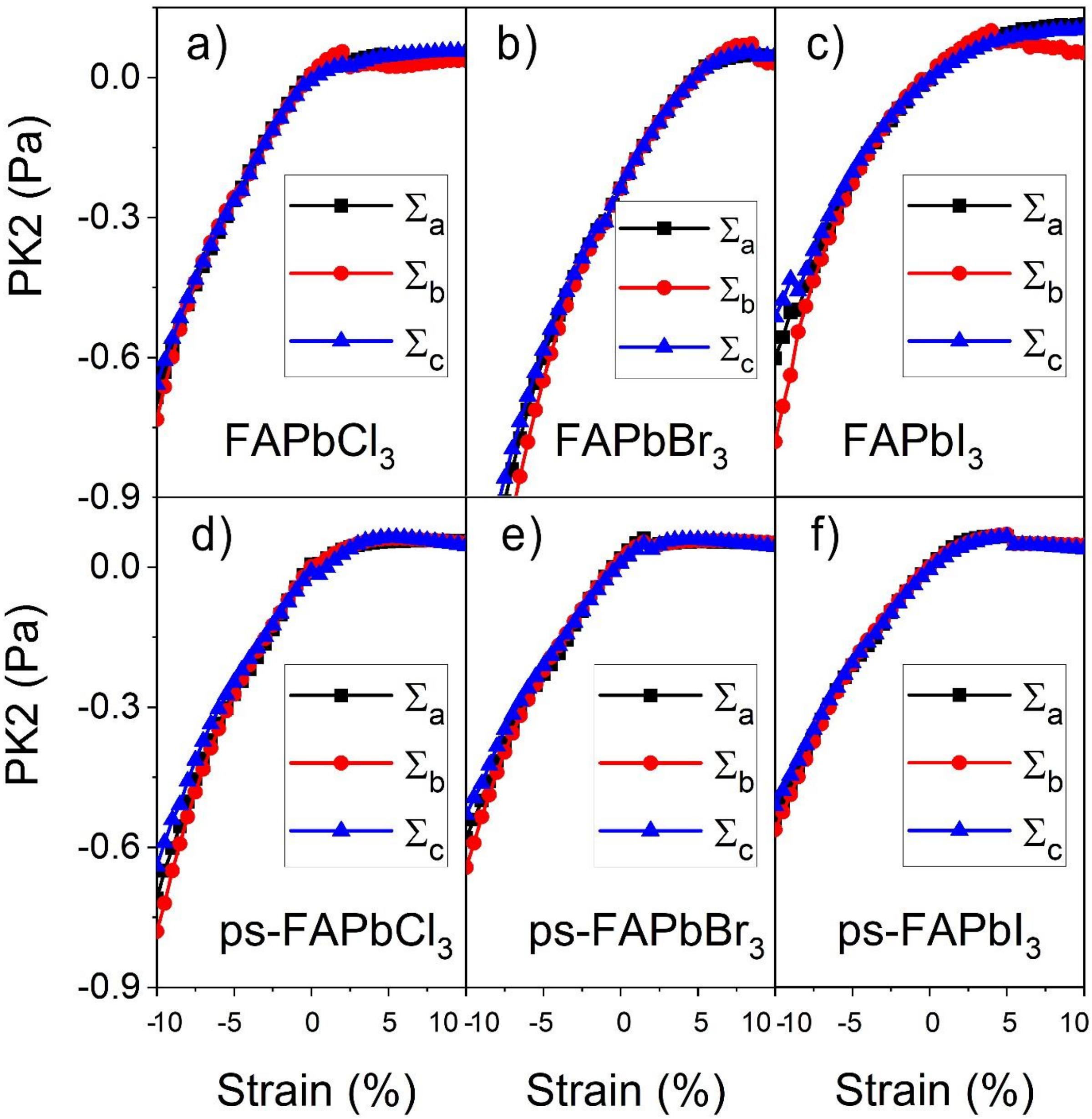

**Figure 3**. Strain responses of $FAPbX_3$ under $D_1$ deformation for the cubic phases of (a) X = Cl, (b) X = Br, and (c) X = I, as well as for the ps-cubic phases of (d) X = Cl, (e) X = Br, and (f) X = I.

These curves (**Figure 3**) enable the definition of the elastic limitation of materials through the concepts of 'ultimate stress' and 'ultimate strain'. The strain corresponding to the highest (ultimate) stress is known as ultimate strain, representing the maximum tension a material can withstand while being stretched. Ultimate stress refers to the highest stress value a material can endure before failure, and ultimate strain indicates the corresponding strain at that point. The elastic limitation of these compounds is determined by the strain range prior to reaching the ultimate strain point. The ultimate stress and strain values for the compounds are listed and compared in **Table 6**. These values were obtained under ideal conditions, neglecting any thermal effects or crystal defects that may influence material behavior in real-world scenarios.

**Table 6**. Ultimate stresses ($\Sigma_u$) and ultimate strain ($\eta_u$) of cubic and ps-cubic $FAPbCl_3$, $FAPbBr_3$ and $FAPbI_3$ in a, b, and c directions under D3 deformation tensor.

| Samples | $FAPbCl_3$ | | $FAPbBr_3$ | | $FAPbI_3$ | |
|---|---|---|---|---|---|---|
| | cubic | ps-cubic | cubic | ps-cubic | cubic | ps-cubic |
| $\Sigma_u^a$ | 0.04 | 0.06 | 0.05 | 0.05 | 0.12 | 0.06 |
| $\eta_u^a$ | 4.5% | 8.5% | 8.5% | 5% | 10% | 4.5% |
| $\Sigma_u^b$ | 0.06 | 0.06 | 0.07 | 0.06 | 0.10 | 0.07 |
| $\eta_u^b$ | 2% | 5.5% | 8.5% | 5.5% | 4% | 5% |
| $\Sigma_u^c$ | 0.06 | 0.06 | 0.05 | 0.06 | 0.10 | 0.06 |
| $\eta_u^c$ | 9% | 5% | 8.5% | 4.5% | 11% | 5% |

The maximum strain that a material can withstand under non-ideal conditions, such as temperature fluctuations and crystal flaws, is always lower than the ultimate strain of a flawless material. The material may experience collapse due to factors such as crystal flaws, vacancies, and thermal fluctuations, as it becomes unstable beyond the maximum strain threshold. Therefore, it is crucial to utilize the data within the ultimate strain zone, where a physical definition exists, for determining the elastic constants. Additionally, it is worth noting that the final strain reflects the intrinsic strength of the structural bonds.

In the case of $FAPbI_3$ and $FAPbBr_3$, the ultimate strain of cubic perovskites is greater than that of ps-cubic perovskites. These findings indicate that while ps-cubic perovskites exhibit greater hardness than cubic perovskites, they are less flexible and undergo rapid shape changes under tensile strains, causing their crystalline structure to collapse. In the cubic structure of $FAPbI_3$, the PK2 diagram at 5% strain exhibits a fracture in the b direction, which gradually decreases with a gentle slope as the strain further increases. However, in the a and c directions, the PK2 diagram remains continuous without any breaks and reaches its maximum at 10% strain. The fracture in the *b* direction corresponds to Pb and I bonds in that direction after reaching 5% strain, causing the $PbI_3$ framework to split into two separate plates (**Figure 3**).

In the case of the cubic $FAPbCl_3$, a small rotation in the FA cation at 2% strain causes the Cl atom to shift from the center of the two Pb atoms along the b direction. Consequently, the initially 2.86 Angstrom-long Cl-Pb bonds transform into two Cl-Pb bonds with lengths of 2.84 and 2.91 Angstroms (**Figure 3**). This change in symmetry results in a break in the PK2 diagram at 2% strain in the b direction. Additionally, at 4.5% strain, the angle of Pb-Cl-Pb bonds shifts from 165.84 degrees to 178.51 degrees, leading to a break in the PK2 diagram in the *a* direction (**Figure 3**). Finally, at 9% strain, the Cl-Pb bonds fracture in the c direction, causing the material to separate (**Figure 3**).

A direct comparison between the cubic and ps-cubic phases reveals that the effect of symmetry reduction strongly depends on the halide species. For $FAPbCl_3$, the ps-cubic phase produces a more isotropic ultimate stress distribution and enhances the mechanical stability along certain crystallographic directions. For $FAPbBr_3$, the cubic phase exhibits greater stability in terms of ultimate strain, whereas the ps-cubic phase displays a more homogeneous but relatively more brittle mechanical response. In the case of $FAPbI_3$, the cubic phase clearly demonstrates the highest stress- and strain-bearing capacity, while the ps-cubic phase enters the softening regime at lower strain because of its initial structural distortion and its closer proximity to mechanically unstable deformation pathways. Therefore, the reduction in symmetry from the cubic to the ps-cubic phase does not necessarily lead to a stiffer material. Instead, depending on the halide composition, it may either activate additional strain-accommodation mechanisms or restrict them, thereby governing the overall mechanical response.

## Conclusions

In this work, the structural, elastic, dynamical, and nonlinear mechanical properties of formamidinium-based halide perovskites, $FAPbX_3$ (X = Cl, Br, and I), were systematically investigated for both the cubic and ps-cubic phases using density functional theory (DFT). The results demonstrate that both phases satisfy the elastic stability criteria; however, the effect of crystal symmetry reduction on the mechanical response is not universal and strongly depends on the halide composition. Analysis of the elastic constants, mechanical moduli, sound velocities, and Debye temperatures reveals that the transition from the cubic to the ps-cubic phase decreases the lattice stiffness and Debye temperature in $FAPbCl_3$ and $FAPbBr_3$, whereas the opposite trend is observed for $FAPbI_3$, where the enhanced elastic moduli and Debye temperature indicate increased lattice rigidity and improved vibrational stability.

The second Piola–Kirchhoff stress–strain analysis further reveals that all compounds exhibit nonlinear, anisotropic, and asymmetric mechanical behavior under both tensile and compressive loading. Moreover, symmetry reduction was found to either activate or suppress strain-accommodation mechanisms depending on the halide species, thereby influencing the ultimate stress and strain, the onset of structural softening, and the overall mechanical stability. Among the investigated compounds, cubic $FAPbI_3$ exhibits the highest stress- and strain-bearing capacity, whereas the mechanical behavior of $FAPbCl_3$ and $FAPbBr_3$ is more sensitive to symmetry reduction.

Overall, this study demonstrates that the mechanical response of formamidinium-based halide perovskites is governed not only by their chemical composition but also by crystal symmetry, which plays a decisive role in controlling lattice stiffness, lattice dynamics, and deformation mechanisms. These findings provide a deeper understanding of the relationship between crystal structure and mechanical properties in this family of perovskites and offer valuable guidance for the design of mechanically robust and reliable optoelectronic materials.

**Acknowledgments**

C. A. was supported by the "MagTop" project (FENG.02.01-IP.05-0028/23) carried out within the "International Research Agendas" programme of the Foundation for Polish Science, co-financed by the European Union under the European Funds for Smart Economy 2021-2027 (FENG). S. M. was supported by the National Science Centre (NCN), Poland, under grant OPUS 21 No. UMO-2021/41/B/ST3/04475. This research was supported by the infrastructure of PCSS (Poznań Supercomputing and Networking Center), grants No. pl0559 and pl0395.

**Supporting information**

# Symmetry-Dependent Mechanical and Vibrational Response of Formamidinium Lead Halide Perovskites: A DFT Study

Mahdi Faghihnasiri[a,*], Carmine Autieri[b] and Sara Memarzadeh[b,c]
[a] Computational Laboratory, Carbon Tech Industrial Group, CarbonTech, Iran
[b] International Research Centre Magtop, Institute of Physics, Polish Academy of Sciences, Aleja Lotników 32/46, 02668 Warsaw, Poland
[c] Institute of Spintronics and Quantum Information, Faculty of Physics and Astronomy, Adam Mickiewicz University Poznań, Uniwersytetu Poznańskiego 2, 61-614 Poznań, Poland

[*] mahdi.faghihnasiri@gmail.com

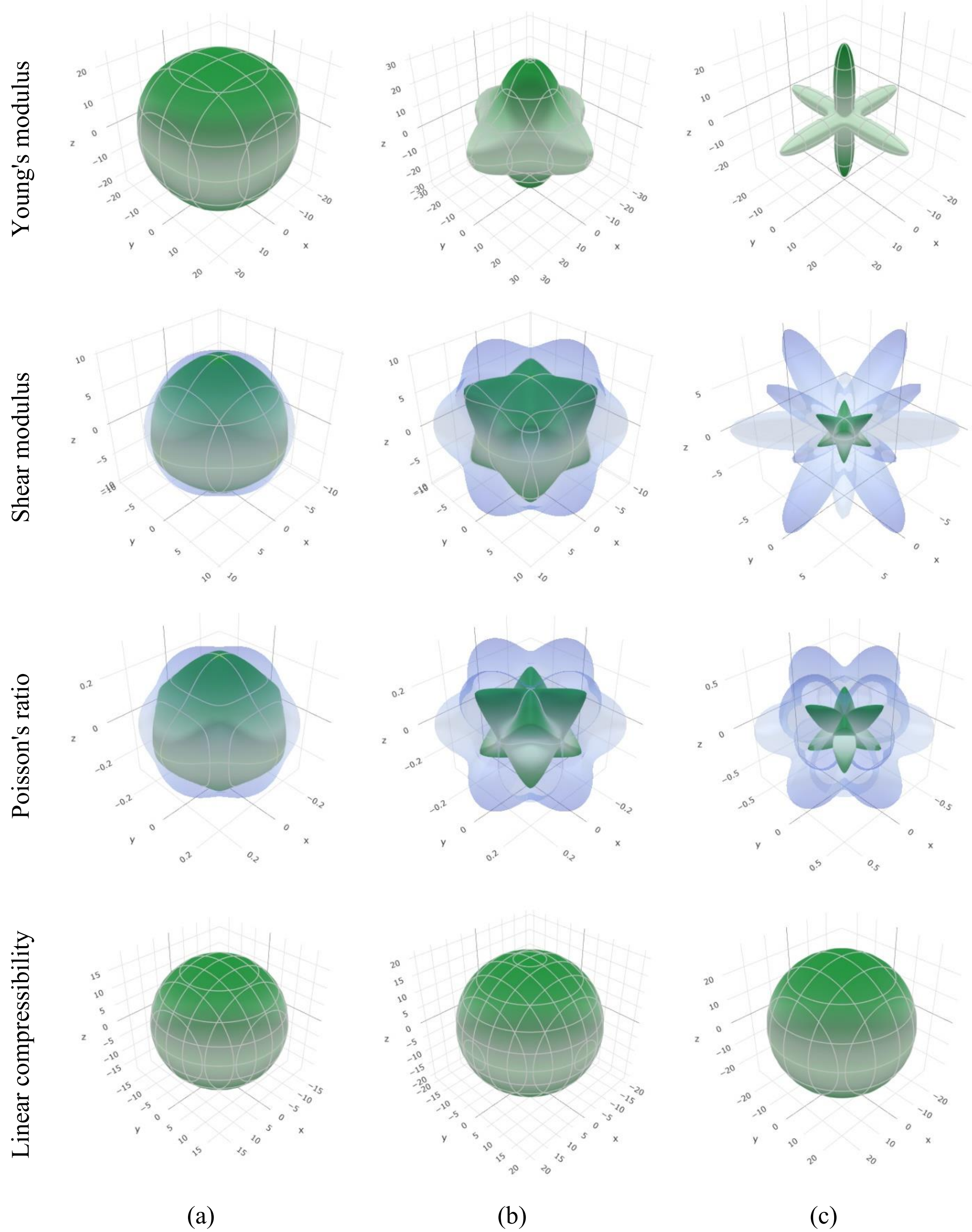

**Figure S1**. 3D contours of Young's, Shear modulus, Linear compressibility and Poisson's ratio for the cubic structure of (a) $FAPbCl_3$, (b) $FAPbBr_3$ and (c) $FAPbI_3$

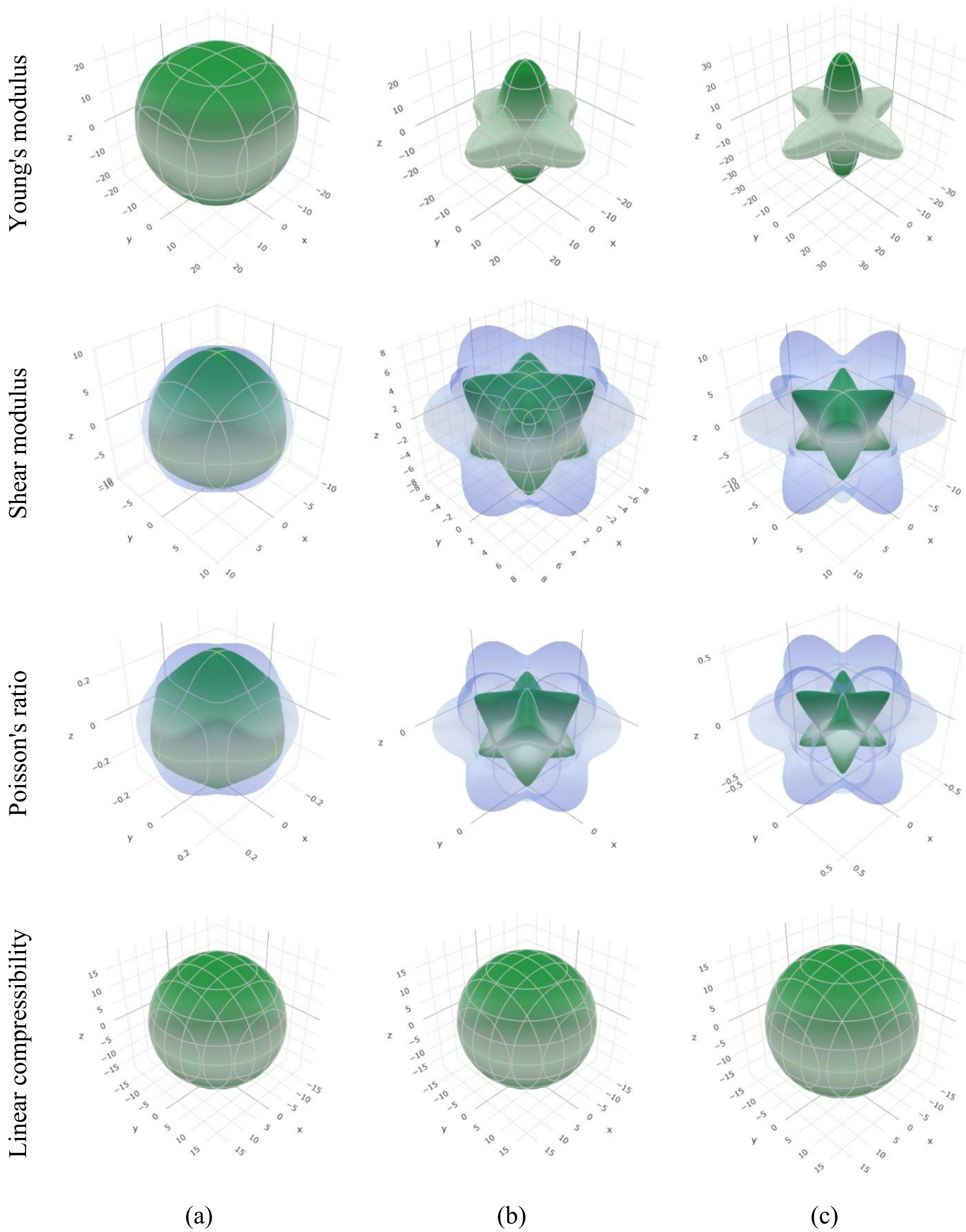

**Figure S2**. 3D contours of Young's, Shear modulus, Linear compressibility and Poisson's ratio for the ps-cubic structure of (a) $FAPbCl_3$, (b) $FAPbBr_3$ and (c) $FAPbI_3$